\documentclass[aps,twocolumn,pra,superscriptaddress,amsmath,showpacs,tightenlines]{revtex4-1}
\usepackage{amssymb}
\usepackage{amsmath}
\usepackage{graphicx}
\usepackage{subfigure}
\usepackage{natbib}
\usepackage{epsfig}
\usepackage{amsfonts}
\usepackage{mathrsfs}
\usepackage{xcolor}
\usepackage[toc,page,title,titletoc,header]{appendix}
\begin{document}

\title{Exact solvability and two-frequency Rabi oscillation in cavity-QED setup with moving emitter}

\author{Mingzhu Weng}
\affiliation{Center for Quantum Sciences and School of Physics, Northeast Normal University, Changchun 130024, China}
\author{Zhihai Wang}
\email{wangzh761@nenu.edu.cn}
\affiliation{Center for Quantum Sciences and School of Physics, Northeast Normal University, Changchun 130024, China}

\begin{abstract}
In this paper, we investigate the energy spectrum and coherent dynamical process in a cavity-QED setup with a moving emitter, which is subject to a harmonic potential. We find that the vibration of the emitter will induce the effective Kerr and optomechanical interactions. We generalize the Bogliubov operators approach which dealt with quantum Rabi model, to our cavity-emitter-vibration system and obtain the energy spectrum exactly. With the assistance of Bogliubov operators approach, we obtain the energy spectrum of the system exactly. Furthermore, we show that the dynamics of the system exhibit a two-frequency Rabi oscillation behavior. We explain such behavior by optomechanical interaction induced quantum transition between emitter-cavity dressed states. We hope that the interaction between cavity mode and moving emitter will provide a versatile platform to explore more exotic effects and potential applications in cavity-QED scenario.
\end{abstract}


\maketitle
\section{introduction}

Light-matter interaction is a fundamental topic in the modern physics, ranging from quantum optics and quantum information processing to the condensed matter physics. The cavity is usually used to adjust the emission of the emitter, leading to the Purcell effect~\cite{purcell1946}, which is a central concept in the field of cavity quantum electrodynamics (QED). In the strong coupling regime, the cavity-QED setup can be described by the Jaynes-Cummings (JC) model~\cite{JC1963}, and the single and multiple photon quantum Rabi oscillation have been studied broadly~\cite{Agarwal1985,Brune1996,Garziano2015}.

In the traditional investigations on cavity-QED and waveguide-QED system, the emitter is usually assumed to be static under the dipole approximation. However, the vibration degrees of freedom of quantum emitters are recently received more and more attentions. For example, in waveguide-QED setup, the waveguide induced interaction between moving emitters has been deeply studied for both of the cases when the velocity of the emitter is faster and slower than that of the photons in the waveguide~\cite{GC2017,ES2020}. Also, the motion of the emitter also leads to the recoil effect~\cite{QL2013,DB2008,FD2016}, which is predicted by the modulated single-photon scattering line shape. Even in the cavity-QED setup, the motion of emitter also induces many interesting phenomena and applications which are absent for the static ones. For example, the oscillation collapse and revival of atomic transition probability~\cite{XG1995, LX2000}, the spatial decoherence~\cite{LZ2005,LZ2010,LY2001}, the motional $n$-phonon bundle state~\cite{YG2021,CS2014}, the exotic photon statistics~\cite{YZ2015,KM2015}, as well as the dynamical Casimir effect~\cite{AA2021,SS2009,OD2019,HW2019,WQ2018,VM2018,SF2015}, just to name a few. On the other hand, in the recent cavity-QED experiment, the Rydberg atom is usually subject to the harmonic potential which is generated by the laser or magneto-optical technology~\cite{Anderson2011,Tikman2016,Bounds2018}. Therefore, it naturally motivates us to investigate the exact energy spectrum and dynamical evolution of the cavity-QED setup with moving emitter which yields to a harmonic potential.

In this work, we focus on the quantum effect of the vibration of the two-level emitter on the energy spectrum and Rabi oscillation of cavity-QED setup. Our model is similar to the trap ion system which is broadly studied to pursuit its application in quantum information processing~\cite{FM2018,FM2012,LD2018,LD2019}. Instead, we here aim to find the exact energy diagram and study the coherent dynamics of the system, in order to achieve a basic understand for the model. With the assistance of a unitary transformation, we find the system can be effectively described by an emitter-optomechanical cavity Hamiltonian~\cite{kippenberg2013,liu2014}, with a negligible Kerr term. We find that effective Hamiltonian possesses a same mathematical structure with quantum Rabi model~\cite{DB2011}, and borrow the Bogliubov operators approach~\cite{QC2012,QH2011,QH2012} to obtain the exact energy spectrum. It shows that, the optomechanical interaction will induce a sideband effect, and in each of the sideband, we observe the Rabi splitting which originates from the emitter-cavity coupling. We also find that the effective optomechanical interaction leads to the two-frequency Rabi oscillation and explain it in the dressed state presentation.

\section{Model and Hamiltonian}
\label{Model}

\begin{figure}
\begin{centering}
\includegraphics[width=1\columnwidth]{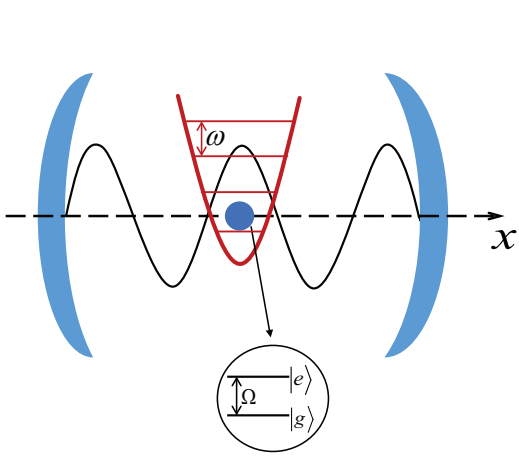}
\par\end{centering}
\caption{Schematic diagram of the model: a single mode cavity couples to a two-level moving emitter which is subject to a harmonic potential.}
\label{device}
\end{figure}
As schematically shown in Fig.~\ref{device}, the system we consider is composed by a single-mode cavity and a movable but spatially confined two-level emitter. The emitter is characterized by its mass $M$ and the internal energy level spacing $\Omega$ between the ground state $|g\rangle $ and excited state $|e\rangle$. We introduce a confinement of the emitter by a harmonic potential of the oscillator frequency $\omega$.
Considering that the spatial motion (vibration) of the emitter is along the $x$ axis, which is perpendicular to the wall of the cavity, the Hamiltonian is given by~\cite{XG1995, LX2000}
\begin{eqnarray}
H&=&\frac{p^{2}}{2M}+\frac{1}{2}M\omega^{2}x^{2}+\hbar\omega_a a^{\dagger}a+\hbar\Omega |e\rangle\langle e|\nonumber \\&&+\hbar g(a^{\dagger}\sigma_- e^{-ikx}+{\rm H.c.}).
\label{h1}
\end{eqnarray}
Here, $x$ and $p$ are the emitter's position and momentum operators, $\omega_a =c k$ is the cavity frequency with $k$ being the photon wave vector, and $c$ being the velocity of light. $\sigma_-=(\sigma_+)^{\dagger}=|g\rangle\langle e|$ is the Pauli operator of the emitter, $a\,({a}^{\dagger})$ is the annihilation (creation) operator of the cavity field. $g$ is the coupling strength between the emitter and cavity. In the Hamiltonian Eq.~(\ref{h1}), we have applied the rotating wave approximation by considering $g\ll\{\omega_a,\Omega\}$.  It is convenient to introduce the creation (annihilation) operator $b^\dagger\,(b)$, which satisfies $x=\alpha (b^{\dagger}+b), p=i\hbar(b^{\dagger}-b)/(2\alpha)$ ($\alpha=\sqrt{\hbar/2M\omega}$) and the Hamiltonian can be rewritten as
\begin{eqnarray}
H&=&\hbar\omega b^{\dagger}b+\hbar\omega_a a^{\dagger}a+\hbar\Omega |e\rangle\langle e|\nonumber \\&&+\hbar g[a^{\dagger}\sigma_- e^{-ik\alpha(b^{\dagger}+b)}+{\rm H.c.}]
\label{h2}
\end{eqnarray}
by neglecting the constant term.

The operator in the exponential term can be eliminated by performing a unitary transformation $\tilde{H}=UHU^{\dagger}$ where  $U=e^{ik\alpha(b^{\dagger}+b)a^{\dagger}a}$, and it yields $\tilde{H}=\tilde{H}_1+\tilde{H}_2$ with
\begin{eqnarray}
\tilde{H}_1
&=&\hbar\chi(a^{\dagger}a)^{2}+\hbar\omega_{a}a^{\dagger}a+\hbar\Omega |e\rangle\langle e|\nonumber \\&&+\hbar g(a^{\dagger}\sigma_{-} +a\sigma_{+})+\hbar\omega b^{\dagger}b,\\
\tilde{H}_2&=&i\hbar\eta a^{\dagger}a(b-b^{\dagger}),
\label{Hamiltonian}
\end{eqnarray}
where
\begin{equation}
\chi= k^{2}\alpha^{2}\omega,\,\eta= k\alpha\omega.
\end{equation}

It is obvious that, the vibrational movement of the emitter induces two effects. The first one is the Kerr effect as shown by the first term of $\tilde{H}_1$, with the strength $\chi=k^2\alpha^2\omega=\hbar k^2/(2M)$, which is independent of oscillator frequency $\omega$. Physically speaking, the movement of the emitter is described by the generation and absorption of the phonon in second quantization representation, and  it is also accompanied by the generation and absorption of photon in the cavity as shown by the emitter-photon interaction Hamiltonian $\hbar g[a^{\dagger}\sigma_- e^{-ik\alpha(b^{\dagger}+b)}+{\rm H.c.}]$. Therefore, it naturally introduces a self-phase modulation to the photon in the cavity. The other one is the effective coupling between the vibrational degree of freedom of the emitter and the cavity mode. As given by $\tilde{H}_2$, it is actually an effective optomechanical interaction~\cite{law1995,marquardt2014} with strength $\eta=k\alpha\omega$, which depends both on the parameters of the emitter and the harmonic potential. Followed by the typical cavity QED system with Rydberg atom, we take $\Omega=\omega_a=10^{5}$\,GHz, $\omega=1\,{\rm GHz}, k=10^{7}\,{\rm m^{-1}}, M=10^{-27}\,{\rm kg}, g=100\,{\rm MHz}$. Within these parameters, we will have $\chi=0.05g$,\,$\eta=g/\sqrt{2}$. Therefore, the strength of the Kerr effect is much weaker than that of the emitter-cavity coupling, that is, $\chi\ll g$.

The above cavity QED model can be experimentally realized in the Rydberg atom platform, in which the parameters can be achieved by $\Omega=\omega_a=10^{5}$\,GHz,  $k=10^{7}\,{\rm m^{-1}}, M=10^{-27}\,{\rm kg}, g=100\,{\rm MHz}$~\cite{Anderson2011,Tikman2016,Bounds2018} . Furhtermore, the trap of the atom can be realized by the optical tweezers technologies and the depth of the harmonica trap $\omega$ can be achieve by hundreds of MHz~\cite{LTC2012,ZY2013,LH2017}. Within these parameters, the strength of the Kerr effect is much weaker than that of the emitter-cavity coupling, that is, $\chi\ll g$.

\begin{figure}
\begin{centering}
\includegraphics[width=1\columnwidth]{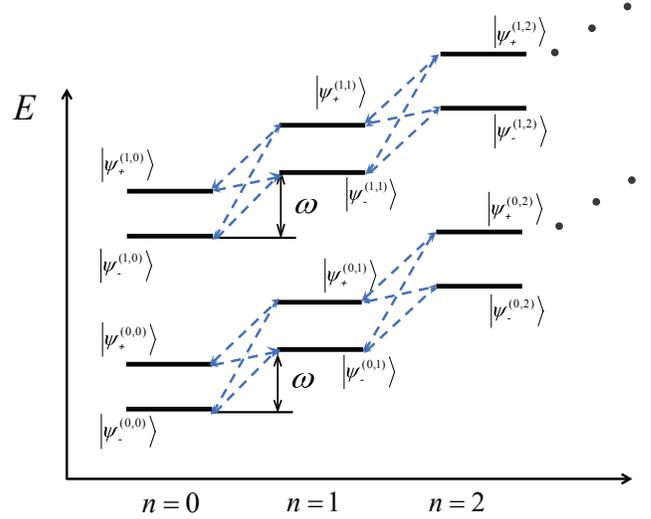}
\par\end{centering}
\caption{The energy spectrum diagram for $m=0,1$. The solid lines are the eigen states of $\tilde{H}_1$ and the dashed lines represent the energy-level transition induced by $\tilde{H}_2$.}
\label{Energyspectrum}
\end{figure}

The Hamiltonian $\tilde{H}_1$ is completely solvable due to the conservation of the excitation number. The eigen values are
\begin{eqnarray}
E_{\pm}^{(m,n)}&=&(m^2+m+\frac{1}{2})\hbar\chi+(m+\frac{1}{2})\hbar\omega_a+\frac{1}{2}\hbar\Omega+n\hbar\omega\nonumber \\
&&\pm\hbar\sqrt{[(m+\frac{1}{2})\chi+\frac{1}{2}\omega_a-\frac{1}{2}\Omega]^2+(m+1)g^2}\nonumber \\
\end{eqnarray}
and the corresponding eigen wave function can be obtained as
\begin{eqnarray}
|\psi_+^{(m,n)}\rangle&=&\cos\frac{\theta}{2}|m,n,e\rangle+\sin\frac{\theta}{2}|m+1,n,g\rangle,\\
|\psi_-^{(m,n)}\rangle&=&-\sin\frac{\theta}{2}|m,n,e\rangle+\cos\frac{\theta}{2}|m+1,n,g\rangle,
\end{eqnarray}
where $\tan\theta=2\sqrt{m+1}g/[\Omega-\omega_{a}-(2m+1)\chi]$, and $|m,n,\sigma\rangle:=|m\rangle_c\otimes|n\rangle_v\otimes|\sigma\rangle_a$ ($|\sigma\rangle=|e\rangle,|g\rangle$) represents the state in which the cavity mode (vibrate mode) is in the bosonic Fock state with $m (n)$ excitations while the emitter is in the state $|\sigma\rangle$.  In Fig.~\ref{Energyspectrum}, we illustrate the energy diagram for $m=0,1$.
Here, the black solid lines are the eigenstates of $\tilde{H}_1$ and the blue dashed lines represent the energy level transitions between $|\psi_{\pm}^{(m,n)}\rangle$ and $|\psi_{\pm}^{(m,n\pm1)}\rangle$, which are induced by $\tilde{H}_2$.   It seems that the whole Hamiltonian can only be solved by the perturbation theory with the presence of $\tilde{H}_2$ induced transition.
However, thanks to the excitation number conservation for the internal degree of freedom for the emitter and the photons in the cavity, that is, $[a^\dagger a+|e\rangle\langle e|,H]=0$, the whole system is still fully solvable, and the exact energy spectrum can be obtained as what we will discuss in the follows.

\section{The solution of the Hamiltonian}
\label{Solution}
Now, we derive the exact energy spectrum of the Hamiltonian $\tilde{H}$. First, we introduce $\tilde{b}=ib$, $\tilde{b^{\dagger}}=-ib^{\dagger}$, then the Hamiltonian $\tilde{H}$ becomes
\begin{eqnarray}
\tilde{H}=&&\hbar\omega\tilde{b^{\dagger}}\tilde{b}+\hbar\omega_{a}a^{\dagger}a+\hbar\Omega |e\rangle\langle e|+\hbar g(a^{\dagger}\sigma_{-} +a\sigma_{+})\nonumber\\
&&+\hbar k\alpha \omega a^{\dagger}a(\tilde{b}+\tilde{b^{\dagger}})
+\hbar k^{2}\alpha^{2}\omega(a^{\dagger}a)^{2}.
\label{H3}
\end{eqnarray}
In what follows, we will still use the symbol $b$ to represent $\tilde{b}$ for the sake of simplicity since it does not affect the final result. In the cavity-emitter basis $\{|m+1,g\rangle,\, |m,e\rangle\}$, the Hamiltonian can be expressed as

\begin{equation}
\tilde{H}=\left( \begin{array}{cc} H_{11} & \hbar\sqrt{m+1} g\\
\hbar \sqrt{m+1} g & H_{22}
\end{array} \right),
\label{Heff}
\end{equation}
where
\begin{eqnarray}
H_{11}&=&\hbar\omega b^{\dagger}b+(m+1)\hbar\omega_a \nonumber \\
&&+(m+1)\hbar k\alpha \omega(b+b^{\dagger})+(m+1)^{2}\hbar k^{2}\alpha^{2}\omega ,\nonumber\\ \\
H_{22}&=&\hbar\omega b^{\dagger}b+m\hbar\omega_a +\hbar\Omega \nonumber \\
&&+m\hbar k\alpha\omega (b+b^{\dagger})+m^{2}\hbar k^{2}\alpha^{2}\omega.
\end{eqnarray}
The Hamiltonian has the same mathematical structure with that of the quantum Rabi model
(see Eq.~(2) in Ref.~\cite{QC2012}). It motivates us to apply the Bogliubov operators approach to solve the eigen spectrum. The basic idea is that we can introduce two Bogolibov transformations to
diagonalize the Hamiltonian $H_{11}$ and $H_{22}$ respectively, and therefore the wave function of the whole Hamiltonian $\tilde{H}$ can be obtained two times. Since they correspond to the same eigenvalue, they should be only different by a complex constant, and then we can build the transcendental equation for the eigen energy. Following the process as given in the appendix (the similar calculation can also be found in Ref.~\cite{QC2012}), the transcendental equation is obtained as $G_m(E)=0$, where
\begin{eqnarray}
G_{0}(E)&=&\sum_{n=0}^{\infty}\left[\frac{g^{2}}{(-n\omega+\gamma+E/\hbar)(\gamma+E/\hbar)}
-1\right]f_{n}(k\alpha)^{n},\nonumber\\
\label{Gfunction0}
\end{eqnarray}
for $m=0$ and
\begin{eqnarray}
G_{m}(E)&=&\sum_{n=0}^{\infty}e_n [k\alpha(m+1)]^{n}\sum_{n=0}^{\infty}e_n (k\alpha m)^{n}\nonumber\\
&&-\sum_{n=0}^{\infty}f_n [k\alpha(m+1)]^{n}\sum_{n=0}^{\infty}f_n (k\alpha m)^{n},
\label{Gfunctionm}
\end{eqnarray}
for $m>0$.

The coefficients $e_n$ and $f_n$ are defined recursively as
\begin{eqnarray}
e_{n}&=&\frac{-\sqrt{m+1}gf_{n}}{l\omega-\gamma-E/\hbar},\\
nf_{n}&=&K_{n-1}f_{n-1}-f_{n-2},
\end{eqnarray}
with the initial conditions $f_{0}=1$ , $f_1 =K_0$, and
\begin{equation}
K_{n}=\frac{1}{k\alpha\omega}[(n\omega+\beta-E/\hbar)-\frac{(m+1)g^{2}}{n\omega-\gamma-E/\hbar}].
\label{Kn}
\end{equation}
Here, $\gamma:=-(m+1)\omega_a,\beta:=k^{2}\alpha^{2}\omega+(m+1)\omega_a$.

\begin{figure}
\begin{centering}
\includegraphics[width=1\columnwidth]{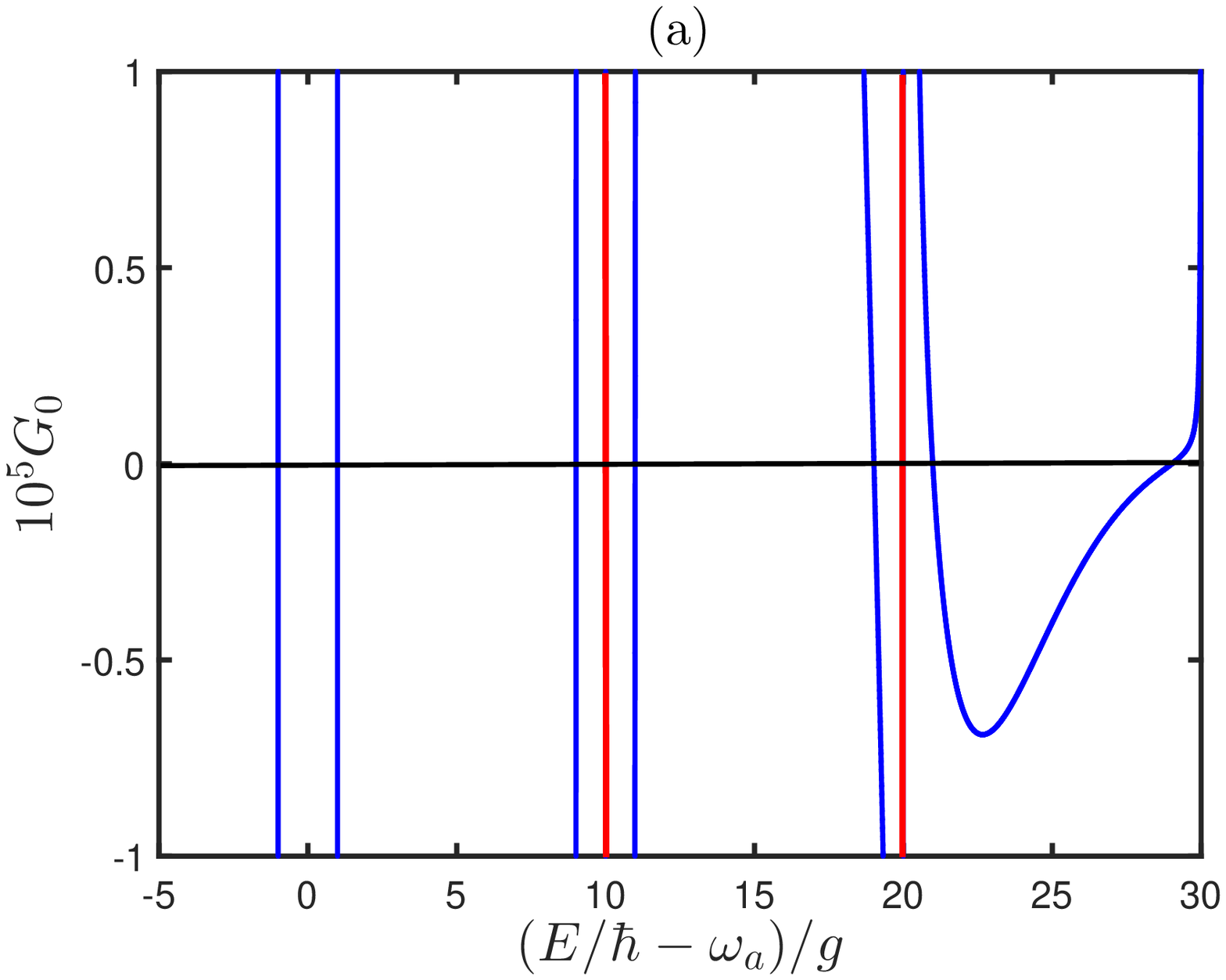}
\includegraphics[width=1\columnwidth]{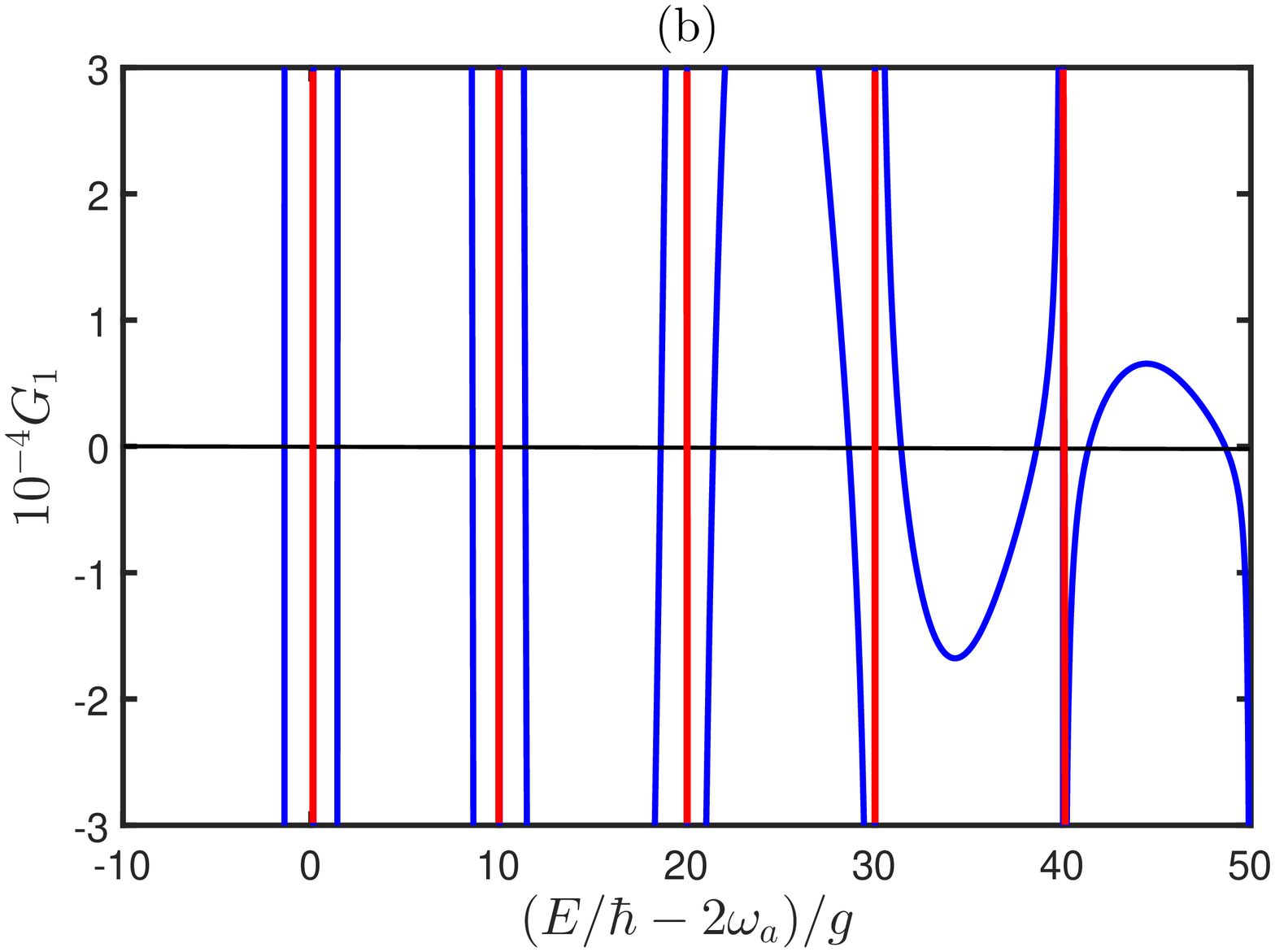}
\end{centering}
\caption{(a) $G_{0}$ and (b) $G_{1}$  for $g=100\,{\rm MHz}, \Omega=\omega_{a}=10^{5}\,{\rm GHz}, k=2\pi/\lambda=10^{7}\,{\rm m^{-1}}$ and $\omega=1\,{\rm GHz}=10g$. }
\label{Gfunction}
\end{figure}

In Fig.~\ref{Gfunction}\,(a) and (b), we plot the $G_m(E)$ functions for $m=0$ and $m=1$ by the blue curves, respectively. Meanwhile, the red curves demonstrate the divergent behavior at $E/\hbar=n\omega+(m+1)\omega_a$, which is implied by Eq.~(\ref{Kn}). Therefore, the zero points of the blue curves correspond to the eigen-energy of the system. As shown in the figure, where we have set the transition frequency being resonant with the cavity, that is $\omega_a=\Omega$, we can clearly observe the sidebands near $n\hbar\omega$, which is induced by the vibration of the emitter. Near each sideband, it shows a Rabi splitting behavior which is given by the emitter-cavity coupling terms $\hbar g(a^\dagger \sigma^-+a\sigma^+)$ in the Hamiltonian.

Recalling that, in Fig.~\ref{Energyspectrum}, we have plotted the eigenstates of $\tilde{H_{1}}$ by the black solid lines\, where the energy level spacing between the states $|\psi_{\pm}^{(m,n)}\rangle$ and $|\psi_{\pm}^{(m,n\pm1)}\rangle$ is
\begin{equation}
\Delta_{m,n}=2\hbar\sqrt{[(m+\frac{1}{2})\chi]^2+(m+1)g^2}.
\end{equation}
For the parameter regime considered in Fig.~\ref{Gfunction}, the energy level space achieves $\Delta_{0,n}\approx2.11\hbar g$ with $m=0$, which is similar to the space $\tilde{\Delta}_{0,n}\approx1.99\hbar g$ in Fig.~\ref{Gfunction}\,(a). The similar result can be also obtained for $m=1$, the result obtained from Eq.~(6) is close to the exact solution given in Fig.~\ref{Gfunction}\,(b) as $|\Delta_{1,n}-\tilde{\Delta}_{1,n}|\approx0.03\hbar g$. Therefore, the energy level transitions introduced by $\tilde{H_{2}}$ produce the slight shift to the energy spectrum of the system.

\section{The Rabi oscillation}
\label{Dynamics}

From now on, we will numerically discuss the dynamical evolution of the system, i.e., to study the Rabi oscillation behavior. Remember that the effective Hamiltonian $\tilde{H}$ is obtained by a unitary transformation, correspondingly, we also need to perform the same unitary transformation on the quantum state.
Therefore, preparing the initial pure state as $|\psi(0)\rangle$, the dynamics of the
system is governed by
\begin{equation}
|\psi(t)\rangle=U^{\dagger}e^{-i\tilde{H}t}U|\psi(0)\rangle,
\end{equation}
and the average value for an arbitrary operator $\hat{A}$ reads
\begin{equation}
\langle \hat{A}\rangle={\rm Tr}[\hat{A}\rho(t)],
\label{generalP}
\end{equation}
where the density matrix $\rho(t)=|\psi(t)\rangle\langle \psi(t)|$.

\begin{figure}
\begin{centering}
\includegraphics[width=1\columnwidth]{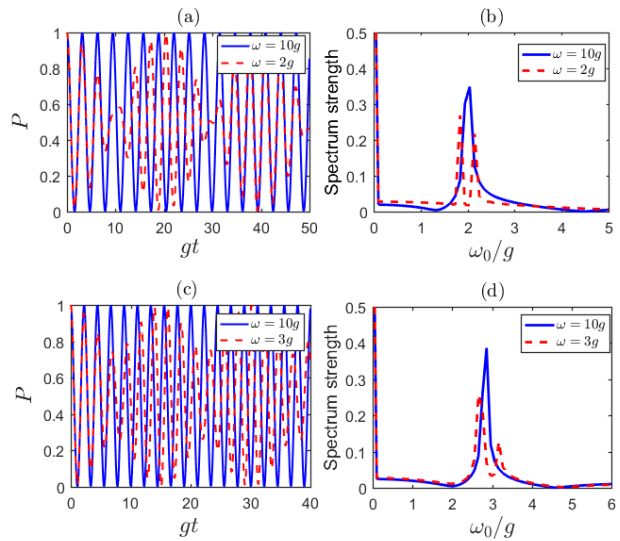}
\par\end{centering}
\caption{The Rabi oscillation of the system (a,c) and the corresponding frequency spectrum (b,d).  The initial state is set as $|\psi(0)\rangle =|m+1,0, g\rangle$ with $m=0$ for (a) and $m=1$ for (c), respectively. The parameters are set to be same as those in Fig.~\ref{Gfunction}.}
\label{Rabi}
\end{figure}

As is well known, for the traditional JC model, the emitter and the cavity field will exchange the excitation, which leads to a perfect Rabi oscillation. However, for a moving emitter, even when the vibrate mode is in the ground state, the oscillation behavior is still changed dramatically.  In Fig.~\ref{Rabi}, we plot  the average value $P=\langle \hat{A}\rangle$ with $\hat{A}=|m,g\rangle\langle m,g|$ with the initial state of the system being $|\psi(0)\rangle =|m+1,0,g\rangle$. In Fig.~\ref{Rabi} (a) and (c), we illustrate the results for $m=0$ and $m=1$, respectively.

As shown in the figure, for a deep harmonic potential $\omega=10g$, the blue curves demonstrate  perfect
Rabi oscillations with periods $T=\pi/g$ and $T=\pi/(\sqrt{2}g)$ respectively in Fig.~\ref{Rabi} (a) and (c). In such a situation, the emitter is confined tightly by the harmonic potential, and it is similar to that in standard JC model with static atom. However, for the shallow harmonic potential, as shown by the red dashed curves in Fig. (a) and (c), the dynamics diverges from the standard Rabi oscillation, and it shows a two-frequency oscillation character, which can be obtained by the numerical Fourier transformation,
\begin{equation}
f(\omega_0)=\frac{1}{\sqrt{2\pi}}\int dt P(t)e^{-i\omega_0t}
\end{equation}
and the spectrum strength corresponds to (a) and (c) are given in (b) and (d), respectively. Here, we clearly observe the spectrum splitting, which is represented by the red dashed lines.

\begin{figure}
\begin{centering}
\includegraphics[width=1\columnwidth]{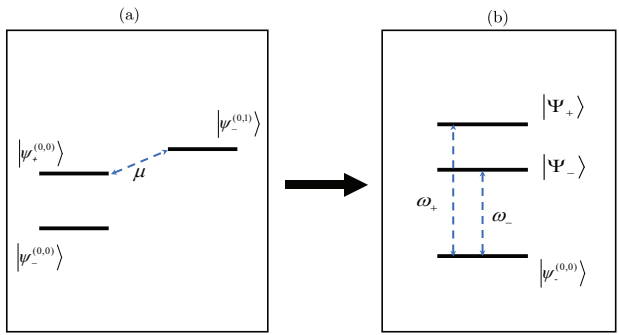}
\par\end{centering}
\caption{(a) The original simplified energy-level diagram. (b) The interpretation for the two-frequency Rabi oscillation  behavior.}
\label{energylevel}
\end{figure}

The splitting can be understood from the energy-level diagram in Fig.~\ref{Energyspectrum}, which shows the energy-level transition between different sidebands. Taking the subspaces with  $m=0$ and $n=0,1$ as an example, the states $|\psi_{\pm}^{(0,0)}\rangle$ will couple to states $|\psi_{\pm}^{(0,1)}\rangle$ simultaneously, that is, it forms four transition channels as shown by the dashed lines. However, in the parameter regime we consider, the coupling between $|\psi_{+}^{(0,0)}\rangle$ and $|\psi_{-}^{(0,1)}\rangle$ will play the most important role due to their smallest energy spacing. As a result, a simplified energy-level diagram can be given by Fig.~\ref{energylevel} (a). Neglecting the effective Kerr interaction, whose strength $\chi$ is much smaller than that of the emitter-cavity coupling $g$, the $|\psi_{+}^{(0,0)}\rangle\leftrightarrow|\psi_{-}^{(0,1)}\rangle$ transition intensity $\mu$ is
\begin{eqnarray}
\hbar \mu\approx\langle\psi_{+}^{(0,0)}|\hbar\eta a^{\dagger}a(b-b^{\dagger})|\psi_{-}^{(0,1)}\rangle
=\frac{1}{2}\hbar\eta.
\end{eqnarray}
with
\begin{eqnarray}
|\psi_{+}^{(0,0)}\rangle&=&\frac{1}{\sqrt{2}}(|0,0,e\rangle+|1,0,g\rangle),\nonumber\\
|\psi_{-}^{(0,1)}\rangle&=&\frac{1}{\sqrt{2}}(-|0,1,e\rangle+|1,1,g\rangle),
\end{eqnarray}
where we have considered $\omega_a=\Omega$. As a result, it forms another two dressed states $|\Psi_+\rangle$ and $|\Psi_-\rangle$, which are the superposition of $|\psi_+^{(0,0)}\rangle$ and $|\psi_-^{(0,1)}\rangle$ as shown in Fig.~\ref{energylevel} (b).  Therefore, the Rabi oscillation process can be approximately considered as the oscillation between the states $|\psi_-^{(0,0)}\rangle$ and $|\Psi_{\pm}\rangle$, the corresponding transition frequency $\omega_{\pm}$ can be obtained as what follows.

Neglecting the Kerr interaction and considering the situation with $\omega_a=\Omega$,  we will have
\begin{subequations}
\begin{eqnarray}
E_+^{(0,0)}&=&\hbar(\omega_a+g),\\
E_-^{(0,1)}&=&\hbar(\omega_a+\omega-g),\\
E_-^{(0,0)}&=&\hbar(\omega_a-g).
\end{eqnarray}
\end{subequations}
Therefore, the eigen energies of $|\Psi_{\pm}\rangle$ are obtained as
\begin{eqnarray}
E_{\pm}&=&\hbar\omega_a+\frac{1}{2}\hbar\omega\pm\hbar\sqrt{(g-\frac{1}{2}\omega)^{2}+\mu^{2}}.
\end{eqnarray}
As a result, the energy level transition frequencies of the system shown in Fig.~\ref{energylevel}(b) are
\begin{eqnarray}
\hbar\omega_{\pm}&=&E_{\pm}-E_-^{(0,0)}\nonumber\\
&=&\hbar g+\frac{1}{2}\hbar\omega\pm\hbar\sqrt{(g-\frac{1}{2}\omega)^{2}+\mu^{2}}.
\end{eqnarray}
For the considered parameters $\omega=2 g$ in Fig.~\ref{Rabi} (a) and (b), the coupling strength achieves $\mu\approx0.15g$, and $\omega_{\pm}\approx (2\pm0.15)g$, which coincides with the two peaks in Fig.~\ref{Rabi} (b) (see the red dashed curve). The similar results can also be obtained for $m=1$, and the two-peak structure for the
spectrum strength which is given by the red dashed curve in Fig.~\ref{Rabi} (d) can be predicted and the transition frequencies are approximately obtained as $\omega_{\pm}'\approx (2.9\pm0.23)g$.

\section{Conclusion}
\label{Conclusion}
In this paper, we investigate the energy spectrum and the Rabi oscillation behavior in a light-matter interaction model with a moving emitter. We introduce a harmonic potential to confine the vibration degree of the emitter, and show that the vibration of the emitter will induce an effective Kerr interaction and opto-mechanical coupling. With the assistance of Bogliubov operators approach, we obtain the exact energy spectrum of the system. Furthermore, with a shallow potential, we find that the Rabi oscillation will exhibit a two-frequency character, which is dramatically different from that of a static emitter.

In the previous studies, it was shown that the mechanical squeezing can be realized in the optomechanical system~\cite{wollman2015}. Therefore, we hope our study about the vibration induced optomechanical interaction can be applied in the squeezed state preparation and is furthermore beneficial for quantum precision measurement and sensing.

{\bf Note-} During the preparation of this work, we find a similar investigation about the optomechanical strong coupling between a single cavity photon and a single atom~\cite{chang2021}.

\begin{acknowledgments}
We thank Profs. X. X. Yi, Y. Li and L.-P. Yang for helpful discussion. This work is supported by the funding from Ministry of Science and Technology of China (No. 2021YFE0193500) and the Natural Science Foundation of China (Nos.~11875011 and
12047566).
\end{acknowledgments}

\appendix
\addcontentsline{toc}{section}{Appendices}\markboth{APPENDICES}{}
\begin{subappendices}
\section{Exact solution}
\label{Appendix}
In this appendix, we give the detailed derivation of the $G$-function, whose zero points yield the eigen energy of the system. The same approach to deal with quantum Rabi model can be found in Ref.~\cite{QC2012}.

Based on Eq.~(\ref{Heff}) in the main text, we introduce the Bogoliubov operators,
\begin{equation}
b^{\dagger}=B^{\dagger}-k\alpha (m+1),\,b=B-k\alpha (m+1),
\end{equation}
to generate the new bosonic operators $B$ and $B^\dagger$. Thus, we can remove the linear term of the diagonal elements of the Hamiltonian matrix and simplify it to
\begin{eqnarray}
\tilde{H}&=&\left( \begin{array}{cc} \hbar\omega B^{\dagger}B-\hbar\gamma & \hbar \sqrt{m+1}g\\
\hbar \sqrt{m+1}g & \hbar\omega B^{\dagger}B-\hbar k\alpha \omega(B^{\dagger}+B)+\hbar\beta
\end{array} \right).\nonumber\\
\end{eqnarray}
where
\begin{equation}
\gamma=-(m+1)\omega_a, \beta=k^{2}\alpha^{2}\omega+m\omega_a+\Omega.
\nonumber
\end{equation}
The wave function can then be assumed as
\begin{eqnarray}
|\Phi\rangle_{B}&=&\left( \begin{array}{c} \sum_{n=0}^{\infty}\sqrt{n!}e_n|n\rangle_B\\
\sum_{n=0}^{\infty}\sqrt{n!}f_n|n\rangle_B
\end{array} \right),
\label{Bn}
\end{eqnarray}
where $e_{n}$ and $f_{n}$ are the expansion coefficients. $|n\rangle_B$ is an extended coherent state. It has the following properties
\begin{eqnarray}
|n\rangle_B&=&\frac{(B^{\dagger})^{n}}{\sqrt{n!}}|0\rangle_B
=\frac{(b^{\dagger}+k\alpha(m+1))^{n}}{\sqrt{n!}}|0\rangle_B,\\
|0\rangle_B&=&e^{-\frac{1}{2}k^{2}\alpha^{2}(m+1)^2-k\alpha(m+1)b^{\dagger}} |0\rangle_b.
\label{B0}
\end{eqnarray}
Here the vacuum state represented by the Bogoliubov operator is defined as the eigenstate of the annihilation operator $b$.

By the Schr\"{o}dinger equation, we will have
\begin{eqnarray}
&&\sum\limits_{n=0}^{\infty}\hbar(n\omega-\gamma)\sqrt{n!}e_{n}|n\rangle_B+\hbar \sqrt{m+1}g\sum\limits_{n=0}^{\infty}\sqrt{n!}f_{n}|n\rangle_B\nonumber\\
&&=E\sum\limits_{n=0}^{\infty}\sqrt{n!}e_{n}|n\rangle_B,\nonumber\\
&&\hbar \sqrt{m+1}g\sum\limits_{n=0}^{\infty}\sqrt{n!}e_{n}|n\rangle_B+\sum\limits_{n=0}^{\infty}\hbar(n\omega+\beta)\sqrt{n!}f_{n}|n\rangle_B\nonumber\\
&&-\hbar k\alpha\omega\sum\limits_{n=0}^{\infty}(\sqrt{n}f_n\sqrt{n!}|n-1\rangle_B+\sqrt{n+1}f_n\sqrt{n!}|n+1\rangle_B)\nonumber\\
&&=E\sum\limits_{n=0}^{\infty}\sqrt{n!}f_{n}|n\rangle_B.\nonumber\\
\end{eqnarray}
Left-multiplying both sides of the above equation by ${}_{B}\langle l|$ gives
\begin{equation}
(l\omega-\gamma-E/\hbar)e_{l}=-\sqrt{m+1}gf_{l},
\end{equation}
\begin{equation}
(l\omega+\beta-E/\hbar)f_{l}-k\alpha\omega(l+1)f_{l+1}-k\alpha\omega=-\sqrt{m+1}ge_{l}.
\end{equation}
The coefficients $e_{l}$ and $f_{l}$ have the following relationship
\begin{eqnarray}
e_{l}&=&\frac{-\sqrt{m+1}gf_{l}}{l\omega-\gamma-E/\hbar}\label{eB},\\
lf_{l}&=&K_{l-1}f_{l-1}-f_{l-2},
\end{eqnarray}
where
\begin{equation}
K(l)=\frac{1}{k\alpha\omega}[(l\omega+\beta-E/\hbar)-\frac{(m+1)g^{2}}{l\omega-\gamma-E/\hbar}].
\end{equation}
with $f_{0}=1$ and $f_{1}=K_0$.

Similarly, we can define another Bogoliubov operator A ($b^{\dagger}=A^{\dagger}-k\alpha m$). The transformed Hamiltonian then reads
\begin{eqnarray}
\tilde{H}&=&\left( \begin{array}{cc} \hbar\omega A^{\dagger}A+\hbar k\alpha \omega(A^{\dagger}+A)+\hbar\beta' & \hbar \sqrt{m+1}g\\
\hbar \sqrt{m+1}g & \hbar\omega A^{\dagger}A-\hbar\gamma'
\end{array} \right),\nonumber\\
\end{eqnarray}
where
\begin{equation}
\gamma'=-m\omega_a-\Omega,\ \beta'=k^{2}\alpha^{2}\omega+(m+1)\omega_a.
\nonumber
\end{equation}
The wave function can also be written as
\begin{eqnarray}
|\Phi\rangle_{A}&=&\left( \begin{array}{c} (-1)^{n}\sum_{n=0}^{\infty}\sqrt{n!}f_{n}'|n\rangle_A\\
(-1)^{n}\sum_{n=0}^{\infty}\sqrt{n!}e_{n}'|n\rangle_A
\end{array} \right),
\label{An}
\end{eqnarray}
and they obey the properties
\begin{eqnarray}
|n\rangle_A&=&\frac{(A^{\dagger})^{n}}{\sqrt{n!}}|0\rangle_A
=\frac{(b^{\dagger}+k\alpha m)^{n}}{\sqrt{n!}}|0\rangle_A,\\
|0\rangle_A&=&e^{-\frac{1}{2}k^{2}\alpha^{2}m^2-k\alpha m b^{\dagger}} |0\rangle_b.
\label{A0}
\end{eqnarray}
Following the previous steps, the relationship between the two coefficients can be obtained as
\begin{eqnarray}
e_{l}'=\frac{-\sqrt{m+1}gf_{l}'}{l\omega-\gamma'-E/\hbar},
\label{eA}
\end{eqnarray}
The corresponding recursive relationship is
\begin{eqnarray}
lf_{l}'&=&K_{l-1}'f_{l-1}'-f_{l-2}',\\
K'(l)&=&\frac{1}{k\alpha\omega}[(l\omega+\beta'-E/\hbar)-\frac{(m+1)g^{2}}{l\omega-\gamma'-E/\hbar}],\nonumber\\
\end{eqnarray}
with $f_{0}'=1$ and $f_1'=K_0'$.

Since both the wave functions (\ref{Bn}) and (\ref{An}) are the true eigenfunction for a nondegenerate eigenvalue $E$, they should be proportional to each other, that is, $|\Phi\rangle_{B}=r|\Phi\rangle_{A}$, where $r$ is a complex constant. Projecting both sides of this identity onto the original vacuum state $_{b}\langle 0|$ , we have
\begin{eqnarray}
\sum_{n=0}^{\infty}\sqrt{n!}e_{n}{}_{b}\langle 0|n\rangle_B&=&r(-1)^{n}\sum_{n=0}^{\infty}\sqrt{n!}f_{n}'{}_{b}\langle 0|n\rangle_A,\nonumber\\
\sum_{n=0}^{\infty}\sqrt{n!}f_{n}{}_{b}\langle 0|n\rangle_B&=&r(-1)^{n}\sum_{n=0}^{\infty}\sqrt{n!}e_{n}'{}_{b}\langle 0|n\rangle_A,\nonumber\\
\end{eqnarray}
and from (\ref{B0}) and (\ref{A0}), we obtain
\begin{eqnarray}
\sqrt{n!}_{b}\langle0|n\rangle_B&=&(k\alpha(m+1))^{n}e^{-\frac{1}{2}k^{2}\alpha^{2}(m+1)^{2}},\nonumber\\
(-1)^{n}\sqrt{n!}_{b}\langle 0|n\rangle_A&=&(-k\alpha m)^{n}e^{-\frac{1}{2}k^{2}\alpha^{2}m^{2}}.
\end{eqnarray}
Then we have to consider the situations with $m=0$ and $m\neq 0$, respectively.

 When $m=0$, eliminating the ratio constant $r$ gives
\begin{equation}
\sum_{n=0}^{\infty}e_n (k\alpha)^{n}\sum_{n=0}^{\infty}e_n'0^{n}=\sum_{n=0}^{\infty}f_n (k\alpha)^{n}\sum_{n=0}^{\infty}f_n'0^{n},
\end{equation}
which yields
\begin{equation}
\sum_{n=0}^{\infty}e_n (k\alpha)^{n}e_{0}'=\sum_{n=0}^{\infty}f_n (k\alpha)^{n}f_{0}'.
\end{equation}
With (\ref{eB}) and (\ref{eA}), we get
\begin{equation}
\sum_{n=0}^{\infty}\frac{-gf_{n}}{n\omega-\gamma-E/\hbar}(k\alpha)^{n}\frac{gf_{0}'}{\gamma'+E/\hbar}-\sum_{n=0}^{\infty}f_n (k\alpha)^{n}f_{0}'=0.
\end{equation}
Setting $\Omega=\omega_a$, we obtain the transcendental equation for the eigen energy $E$ as
\begin{eqnarray}
G_{0}(E)&=&\sum_{n=0}^{\infty}\frac{g^{2}f_{n}}{(-n\omega+\gamma+E/\hbar)(\gamma+E/\hbar)}(k\alpha)^{n}\nonumber\\
&&-\sum_{n=0}^{\infty}f_n (k\alpha)^{n}=0.
\end{eqnarray}
For $m\neq 0$, we will similarly reach
\begin{eqnarray}
G_{m}(E)&=&\sum_{n=0}^{\infty}e_n [k\alpha(m+1)]^{n}\sum_{n=0}^{\infty}e_n (k\alpha m)^{n}\nonumber\\
&&-\sum_{n=0}^{\infty}f_n [k\alpha(m+1)]^{n}\sum_{n=0}^{\infty}f_n (k\alpha m)^{n}=0.\nonumber\\
\end{eqnarray}
which are Eq.~(\ref{Gfunction0}) and Eq.~(\ref{Gfunctionm}) in the main text for $m=0$ and $m\neq0$, respectvely.
\end{subappendices}


\begin{thebibliography}{99}

\bibitem{purcell1946} E. M. Purcell, Phys. Rev. {\bf 69}, 681 (1946).

\bibitem{JC1963}E. T. Jaynes and F. W. Cummings, Proc. IEEE {\bf 51}, 89 (1963).

\bibitem{Agarwal1985} G. S. Agarwal,  J. Opt. Soc. Am. B {\bf 2}, 480 (1985)

\bibitem{Brune1996} M. Brune, F. S.-Kaler, A. Maali, J. Dreyer, E. Hagley,
J. M. Raimond, and S. Haroche,  Phys. Rev. Lett. {\bf 76}, 1800 (1996).

\bibitem{Garziano2015}L. Garziano, R. Stassi, V. Macr\'{\i}, A. F. Kockum, S. Savasta and F. Nori,
Phys. Rev. A {\bf 92}, 063830 (2015).


\bibitem{GC2017}G. Calaj\'{o} and P. Rabl, Phys. Rev. A {\bf 95}, 043824 (2017).

\bibitem{ES2020}E. S.-Burillo, A. G.-Tudela, and C. G.-Ballestero, Phys. Rev. A {\bf 102}, 013726 (2020).


\bibitem{QL2013}Q. Li, D. Z. Xu, C. Y. Cai, and C. P. Sun, Sci. Rep. {\bf 3}, 3144 (2013).

\bibitem{DB2008}D. Braun and J. Martin, Phys. Rev. A {\bf 77}, 032102 (2008).

\bibitem{FD2016}F. Damanet, D. Braun, and J. Martin, Phys. Rev. A {\bf 93}, 022124 (2016).

\bibitem{XG1995}X. G. Wang and C. P. Sun, J. Mod. Optics {\bf 42}, 515 (1995).

\bibitem{LX2000}L. X. Cen and S. J. Wang, J. Phys. A: Math. Gen. {\bf 33}, 3697 (2000).


\bibitem{LZ2005}L. Zheng, C. Li, Y. Li, and C. P. Sun, Phys. Rev. A {\bf 71}, 062101 (2005).

\bibitem{LZ2010}L. Zheng, C. P. Yang, and F. Nori, Phys. Rev. A {\bf 82}, 062106 (2010).

\bibitem{LY2001}L. You, Phys. Rev. A {\bf 64}, 012302 (2001).


\bibitem{YG2021}Y. G. Deng, T. Shi, and S. Yi, Photon. Res. {\bf 9}, 1289 (2021).

\bibitem{CS2014}C. S. Mu\~{n}oz, E. d. Valle, A. G. Tudela, K. M\"{u}ller, S. Lichtmannecker, M. Kaniber, C. Tejedor, J. J. Finley, and F. P. Laussy,  Nat. Photon. {\bf 8}, 550 (2014).


\bibitem{YZ2015}Y. Zhang, J. Zhang, S. X. Wu, and C. S. Yu, Ann. Phys. {\bf 361}, 563 (2015).

\bibitem{KM2015}K. M. Birnbaum, A. Boca, R. Miller, A. D. Boozer, T. E. Northup, and H. J. Kimble, Nature {\bf 436}, 87 (2005).



\bibitem{AA2021}A. Agust\'{i}, L. G. \'{A}lvarez, E. Solano, and C. Sab\'{i}n, Phys. Rev. A {\bf 103}, 062201 (2021).

\bibitem{SS2009}S. Scheel and S. Y. Buhmann, Phys. Rev. A {\bf 80}, 042902 (2009).

\bibitem{OD2019}O. D. Stefano, A. Settineri, V. Macr\'{i}, A. Ridolfo, R. Stassi, A. F. Kockum, S. Savasta, and F. Nori, Phys. Rev. Lett. {\bf 112}, 030402 (2019).

\bibitem{HW2019}H. Wang, M. P. Blencowe, C. M. Wilson, and A. J. Rimberg, Phys. Rev. A {\bf 99}, 053833 (2019).

\bibitem{WQ2018}W. Qin, V. Macr\'{i}, A. Miranowicz, S. Savasta, and F. Nori, Phys. Rev. A {\bf 100}, 062501 (2019).

\bibitem{VM2018}V. Macr\'{i}, A. Ridolfo, O. D. Stefano, A. F. Kockum, F. Nori, and S. Savasta, Phys. Rev. X {\bf 8}, 011031 (2018).

\bibitem{SF2015}S. Felicetti, C. Sab\'{i}n, I. Fuentes, L. Lamata, G. Romero, and E. Solano, Phys. Rev. B {\bf 92}, 064501 (2015).

\bibitem{Anderson2011}S. E. Anderson, K. C. Younge, and G. Raithel, Phys. Rev. Lett. {\bf 107}, 263001 (2011).

\bibitem{Tikman2016}Y. Tikman, I. Yavuz, M. F. Ciappina, A. Chac\'{o}n, Z. Altun, and M. Lewenstein, Phys. Rev. A {\bf 93}, 023410 (2016).

\bibitem{Bounds2018}A. D. Bounds, N. C. Jackson, R. K. Hanley, R. Faoro, E. M. Bridge, P. Huillery, and M. P. A. Jones, Phys. Rev. Lett. {\bf 120}, 183401 (2018).

\bibitem{FM2018}F. Zhou, L. Yan, S. Gong, Z. Ma, J. He, T. Xiong, L. Chen, W. Yang, M. Feng and V. Vedral, Sci. Adv., {\bf 2}, e1600578 (2018).

\bibitem{FM2012}L. Chen, W. Wan, Y. Xie, F. Zhou and M. Feng, Chin. Phys. Lett., {\bf 29}, 033701 (2012).

\bibitem{LD2018}C. J. Trout, M. Li, M. Guti\'{e}rrez, Y. Wu, S.-T. Wang, L. Duan and K. R. Brown, New J. Phys., {\bf 20}, 043038 (2018).

\bibitem{LD2019}K. A. Landsman, Y. Wu, P. H. Leung, D. Zhu, N. M. Linke, K. R. Brown, L. Duan and C. Monroe, Phys. Rev. A, {\bf 100}, 022332 (2019).

\bibitem{kippenberg2013}T. Ramos, V. Sudhir, K. Stannigel, P. Zoller, and T. J. Kippenberg, Phys. Rev. Lett. {\bf 110}, 193602 (2013).

\bibitem{liu2014}H. Wang, X. Gu, Y.-x. Liu, A. Miranowicz, and F. Nori, Phys. Rev. A {\bf 90}, 023817 (2014).

\bibitem{DB2011}D. Braak, Phys. Rev. Lett. {\bf 107}, 100401 (2011).

\bibitem{QC2012}Q. H. Chen, C. Wang, S. He, T. Liu, and K. Wang, Phys. Rev. A {\bf 86}, 023822 (2012).

\bibitem{QH2011}Q. H. Chen, T. Liu, Y. Y. Zhang, and K. L. Wang, Europhys. Lett. {\bf 96}, 14003 (2011).

\bibitem{QH2012}Q. H. Chen, L. Li, T. Liu, and K. L. Wang, Chin. Phys. Lett. {\bf 29}, 014208 (2012).

\bibitem{LTC2012}T. Li, Z.-X. Gong, Z.-Q. Yin, H. T. Quan, X. Yin, P. Zhang, L.-M. Duan and X. Zhang, Phys. Rev. Lett., {\bf 109}, 163001 (2012).

\bibitem{ZY2013}Z.-Q. Yin, T. Li, X. Zhang and L.M.  Duan, Phys. Rev. A, {\bf 88}, 033614 (2013) .

\bibitem{LH2017}H.-K. Li, E. Urban, C. Noel, A. Chuang, Y. Xia, A. Ransford, B. Hemmerling, Y. Wang, T. Li, H. H\"{a}ffner and  X. Zhang, Phys. Rev. Lett., {\bf 118}, 053001 (2017).

\bibitem{law1995}C. K. Law, Phys. Rev. A {\bf 51}, 2537 (1995).

\bibitem{marquardt2014} M. Aspelmeyer, T. J. Kippenberg, and F. Marquardt,
Rev. Mod. Phys. {\bf 86}, 1391 (2014).

\bibitem{wollman2015}E. E. Wollman, C. U. Lei, A. J. Weinstein, J. Suh, A. Kronwald, F.
Marquardt, A. A. Clerk, and K. C. Schwab, Science {\bf 349}, 952 (2015).


\bibitem{chang2021} J. A.-Luengo and D. E. Chang, arXiv: 2108.03526 (2021).

\end{thebibliography}
\end{document}